\documentclass[letterpaper,twocolumn,showpacs,prl,aps,superscriptaddress,floatfix]{revtex4}
\usepackage[latin1]{inputenc}
\usepackage{bm}
\usepackage{multirow,amssymb,amsbsy,amsmath}
\usepackage{graphicx}
\makeatletter
\usepackage{pifont}
\makeatother

\begin{document}

\title{Measure for the degree of non-Markovian behavior of quantum processes in open systems}

\author{Heinz-Peter Breuer}

\email{breuer@physik.uni-freiburg.de}

\affiliation{Physikalisches Institut, Universit\"at Freiburg,
Hermann-Herder-Strasse 3, D-79104 Freiburg, Germany}

\author{Elsi-Mari Laine}

\affiliation{Turku Center for Quantum Physics, Department of
Physics and Astronomy, University of Turku, FI-20014 Turun
yliopisto, Finland}

\author{Jyrki Piilo}

\email{jyrki.piilo@utu.fi}

\affiliation{Turku Center for Quantum Physics, Department of
Physics and Astronomy, University of Turku, FI-20014 Turun
yliopisto, Finland}

\date{\today}

\begin{abstract}
We construct a general measure for the degree of non-Markovian
behavior in open quantum systems. This measure is based on the
trace distance which quantifies the distinguishability of quantum
states. It represents a functional of the dynamical map describing
the time evolution of physical states, and can be interpreted in
terms of the information flow between the open system and its
environment. The measure takes on nonzero values whenever there is
a flow of information from the environment back to the open
system, which is the key feature of non-Markovian dynamics.
\end{abstract}


\maketitle

The prototype of a Markov process in an open quantum system is
given by a quantum dynamical semigroup, i.e., by the solutions of
a master equation for the reduced density matrix with Lindblad
structure \cite{GORINI,LINDBLAD}. However, in realistic physical
systems the assumption of a Markovian dynamics can only be an
approximation that relies on a number of mostly rather drastic
simplifications. In complex quantum systems one therefore often
encounters dynamical processes which deviate not only
quantitatively but also qualitatively from the relatively simple
behavior predicted by a Markovian time evolution
\cite{BREUER2007}.

In view of the large variety of conceptually different analytical
methods and numerical simulation techniques that have been
developed to treat non-Markovian systems in recent years (see, for
example,
Refs.~\cite{Piilo2007,EISI,HPB08,BREUER2004a,PASCAZIO,LIDAR,LENDI,BUDINI,BARNETT,WILKIE,CRESSER,KOSSAKOWSKI}),
the following questions arise: How can one rigorously define
quantum non-Markovianity and how can one quantify the degree of
non-Markovian behavior in a way which does not refer to any
specific representation or approximation of the dynamics, e.g. to
a master equation with a given structure? In order to answer these
questions one needs a measure for the non-Markovianity of the
quantum dynamics of open systems which, in mathematical terms,
represents a functional of the dynamical map that describes the
time evolution of physical states.

Here, we construct such a measure for non-Markovianity. This
measure is based on the trace distance of two quantum states which
describes the probability of successfully distinguishing these
states. The basic idea underlying our construction is that
Markovian processes tend to continuously reduce the
distinguishability between any two states, while the essential
property of non-Markovian behavior is the growth of this
distinguishability. Interpreting the loss of distinguishabilty of
states as a flow of information from the open system to its
environment, one is thus led to a simple, intuitive picture,
namely that the key feature of non-Markovian dynamics is a
reversed flow of information from the environment back to the open
system. An important consequence of this picture is that the
dynamical map of non-Markovian processes must necessarily be
non-divisible, a property that is known to play also a decisive
role in the classification of quantum channels \cite{WOLF}.

To construct the measure for non-Markovianity we first need a
measure for the distance of two quantum states $\rho_1$ and
$\rho_2$. Such a measure is given by the trace distance (see,
e.g., Ref.~\cite{NIELSEN}) which is defined by
\begin{equation}
 D(\rho_1,\rho_2) = \frac{1}{2}{\mathrm{tr}}|\rho_1-\rho_2|,
\end{equation}
where $|A|=\sqrt{A^{\dagger}A}$. The trace distance $D$ represents
a natural metric on the space of density matrices, i.e., on the
space of physical states, satisfying $0\leq D \leq 1$. Besides
many other interesting properties, it has a clear physical
interpretation in terms of the distinguishability of quantum
states. Suppose that Alice prepares a quantum system in one of two
states $\rho_1$ and $\rho_2$, each with probability $\frac{1}{2}$,
and gives the system to Bob who performs a measurement to decide
whether the system was in the state $\rho_1$ or $\rho_2$. One can
show that the quantity $\frac{1}{2}[1+D(\rho_1,\rho_2)]$ is then
equal to the probability that Bob can successfully identify the
state of the system. Thus, the trace distance can be interpreted
as a measure for the distinguishability of two quantum states. A
further remarkable feature of the trace distance is given by the
fact that all completely positive and trace preserving (CPT) maps
$\Phi$ are contractions for this metric \cite{RUSKAI},
\begin{equation} \label{contraction-1}
 D(\Phi\rho_1,\Phi\rho_2) \leq D(\rho_1,\rho_2).
\end{equation}
This means that no trace preserving quantum operation can ever
increase the distinguishability of two states.

Suppose now that we have a quantum process given by a Markovian
master equation,
\begin{equation} \label{QMEQ-1}
 \frac{d}{dt}\rho(t) = {\mathcal{L}}\rho(t),
\end{equation}
with a generator in Lindblad form \cite{GORINI,LINDBLAD},
\begin{equation} \label{LINDBLAD-GEN}
 {\mathcal{L}}\rho = -i[H,\rho] + \sum_i \gamma_i\left[ A_i\rho A^{\dagger}_i
 - \frac{1}{2}\left\{A^{\dagger}_iA_i,\rho\right\} \right],
\end{equation}
involving a time-independent Hamiltonian $H$ as well as
time-independent Lindblad operators $A_i$ and positive relaxation
rates $\gamma_i\geq 0$. Such a master equation leads to a
dynamical semigroup of CPT maps $\Phi(t) = \exp({\mathcal{L}}t)$,
$t \geq 0$, which describes the dynamics of the density matrix
through the relation $\rho(t)=\Phi(t)\rho(0)$. By use of the
semigroup property $\Phi(\tau+t)=\Phi(\tau)\Phi(t)$ it easily
follows from Eq.~(\ref{contraction-1}) that for all $\tau,t\geq 0$
we have
\begin{equation} \label{contraction-2}
 D(\rho_1(\tau+t),\rho_2(\tau+t)) \leq D(\rho_1(t),\rho_2(t)),
\end{equation}
where $\rho_{1,2}(t)=\Phi(t)\rho_{1,2}(0)$. Thus, for all quantum
dynamical semigroups $\Phi(t)$ the trace distance of the states
$\rho_{1,2}(t)$, corresponding to any fixed pair of initial states
$\rho_{1,2}(0)$, is a monotonically decreasing function of time.
This is a general feature of quantum Markov processes, implying
that under a Markovian evolution any two initial states generally
become less and less distinguishable as time increases. We can
interpret this loss of distinguishability as a certain flow of
information from the system to the environment which continuously
reduces our ability to distinguish the given states.

The inequality (\ref{contraction-2}) holds for a much larger class
of quantum processes than those described by a master equation of
the form (\ref{QMEQ-1}). In fact, suppose we have a time-local
master equation of the form
\begin{equation} \label{QMEQ-2}
 \frac{d}{dt}\rho(t) = {\mathcal{K}}(t)\rho(t)
\end{equation}
with a time-dependent generator ${\mathcal{K}}(t)$. One can show
that in order to preserve the Hermiticity and trace of the density
matrix this generator must be of the form
\cite{GORINI,BREUER2004a}
\begin{eqnarray} \label{TCL-GENERATOR}
 {\mathcal{K}}(t)\rho &=& -i[H(t),\rho] \\
 &~& + \sum_i \gamma_i(t)\left[ A_i(t)\rho A^{\dagger}_i(t)
 - \frac{1}{2}\left\{A^{\dagger}_i(t)A_i(t),\rho\right\} \right],
 \nonumber
\end{eqnarray}
where the Hamiltonian $H(t)$, the Lindblad operators $A_i(t)$ and
the relaxation rates $\gamma_i(t)$ depend on time. If the
relaxation rates are positive functions, $\gamma_i(t)\geq 0$, the
generator (\ref{TCL-GENERATOR}) is seen to be in the Lindblad form
(\ref{LINDBLAD-GEN}) for each fixed $t\geq 0$. Such processes with
$\gamma_i(t)\geq 0$ may be called time-dependent Markovian
although the corresponding dynamical maps $\Phi(t)$ do not lead to
a quantum dynamical semigroup. With the help of the chronological
time-ordering operator ${\mathrm{T}}$ we can define a
two-parameter family of CPT maps $\Phi(t_2,t_1)$ by means of
\begin{equation} \label{TWO-FAMILY}
 \Phi(t_2,t_1) = {\mathrm{T}}\exp\left[ \int_{t_1}^{t_2} dt' {\mathcal{K}}(t')
 \right].
\end{equation}
The dynamical map which transforms the initial states at time $0$
into the states at time $t$ can then be written as
$\Phi(t)=\Phi(t,0)$. The important point to note is that this
dynamical map has the property of being divisible in the sense
that for all $\tau,t\geq 0$ the CPT map $\Phi(\tau+t,0)$ can be
written as composition of the two CPT maps $\Phi(\tau+t,t)$ and
$\Phi(t,0)$,
\begin{equation} \label{DIVISIBILITY}
 \Phi(\tau+t,0) = \Phi(\tau+t,t)\Phi(t,0).
\end{equation}
We remark that for a dynamical semigroup on has
$\Phi(t_2,t_1)=\Phi(t_2-t_1)$ such that Eq.~(\ref{DIVISIBILITY})
reduces to $\Phi(\tau+t)=\Phi(\tau)\Phi(t)$. Since in
Eq.~(\ref{DIVISIBILITY}) not only $\Phi(\tau+t,0)$ and $\Phi(t,0)$
but also $\Phi(\tau+t,t)$ is a CPT map, we conclude that the
relation (\ref{contraction-2}) holds true for all time-dependent
Markovian quantum processes defined by the master equation
(\ref{QMEQ-2}) with $\gamma_i(t)\geq 0$.

We define the rate of change of the trace distance by
\begin{equation} \label{SIGMA}
 \sigma(t,\rho_{1,2}(0)) = \frac{d}{dt}D(\rho_1(t),\rho_2(t)).
\end{equation}
For a given quantum process $\Phi(t)$, this quantity depends on
time $t$ and on the initial states $\rho_{1,2}(0)$ with
corresponding time-evolutions $\rho_{1,2}(t) =
\Phi(t,0)\rho_{1,2}(0)$. As has been demonstrated above, we have
$\sigma\leq 0$ for all quantum processes for which the
divisibility property (\ref{DIVISIBILITY}) holds, i.e., for all
dynamical semigroups and all time-dependent Markovian processes.
We remark that Eq.~(\ref{contraction-1}) not only holds for CPT
maps, but also for the larger class of positive and
trace-preserving maps \cite{RUSKAI}. Thus, $\sigma\leq 0$ holds
true also for Markovian master equations which are not in Lindlbad
form but preserve positivity.

There are however many physical processes for which $\sigma$ is
larger than zero for certain times. It is this type of processes
which we define as non-Markovian. Hence, a process is said to be
non-Markovian if there exists a pair of initial states
$\rho_{1,2}(0)$ and a certain time $t$ such that
$\sigma(t,\rho_{1,2}(0))> 0$. Physically, this means that for
non-Markovian dynamics the distinguishability of the pair of
states increases at certain times. We interpret this as a flow of
information from the environment back to the system which enhances
the possibility of distinguishing the two states. While Markovian
processes tend to wash out more and more characteristic features
of the two states, non-Markovian processes lead to an uncovering
of these features. We emphasize that the temporary backflow of
information represents a natural feature occurring in many
physical systems which does not imply that there is no
thermalization for long times.

How can one construct a measure for non-Markovianity on the basis
of this definition? Clearly, such a quantity should measure the
total increase of the distinguishability over the whole
time-evolution, i.e., the total amount of information which flows
from the environment back to the system. This suggests defining a
measure ${\mathcal{N}}(\Phi)$ for the non-Markovianity of the
quantum process $\Phi(t)$ by means of the relation
\begin{equation} \label{MEASURE-1}
 {\mathcal{N}}(\Phi) = \max_{\rho_{1,2}(0)} \int_{\sigma > 0}
 dt \; \sigma(t,\rho_{1,2}(0)).
\end{equation}
Here, the time-integration is extended over all time intervals
$(a_i,b_i)$ in which $\sigma$ is positive, and the maximum is
taken over all pairs of initial states. In view of
Eq.~(\ref{SIGMA}) we can thus write this definition as
\begin{equation} \label{MEASURE-2}
 {\mathcal{N}}(\Phi) = \max_{\rho_{1,2}(0)} \sum_i \Big[
 D(\rho_1(b_i),\rho_2(b_i)) - D(\rho_1(a_i),\rho_2(a_i))
 \Big].
\end{equation}
To calculate this quantity one first determines for any pair of
initial states the total growth of the trace distance over each
time interval $(a_i,b_i)$ and sums up the contributions of all
intervals. ${\mathcal{N}}(\Phi)$ is then obtained by determining
the maximum over all pairs of initial states.

\begin{figure}[htb]
\begin{center}
\includegraphics[scale=0.34]{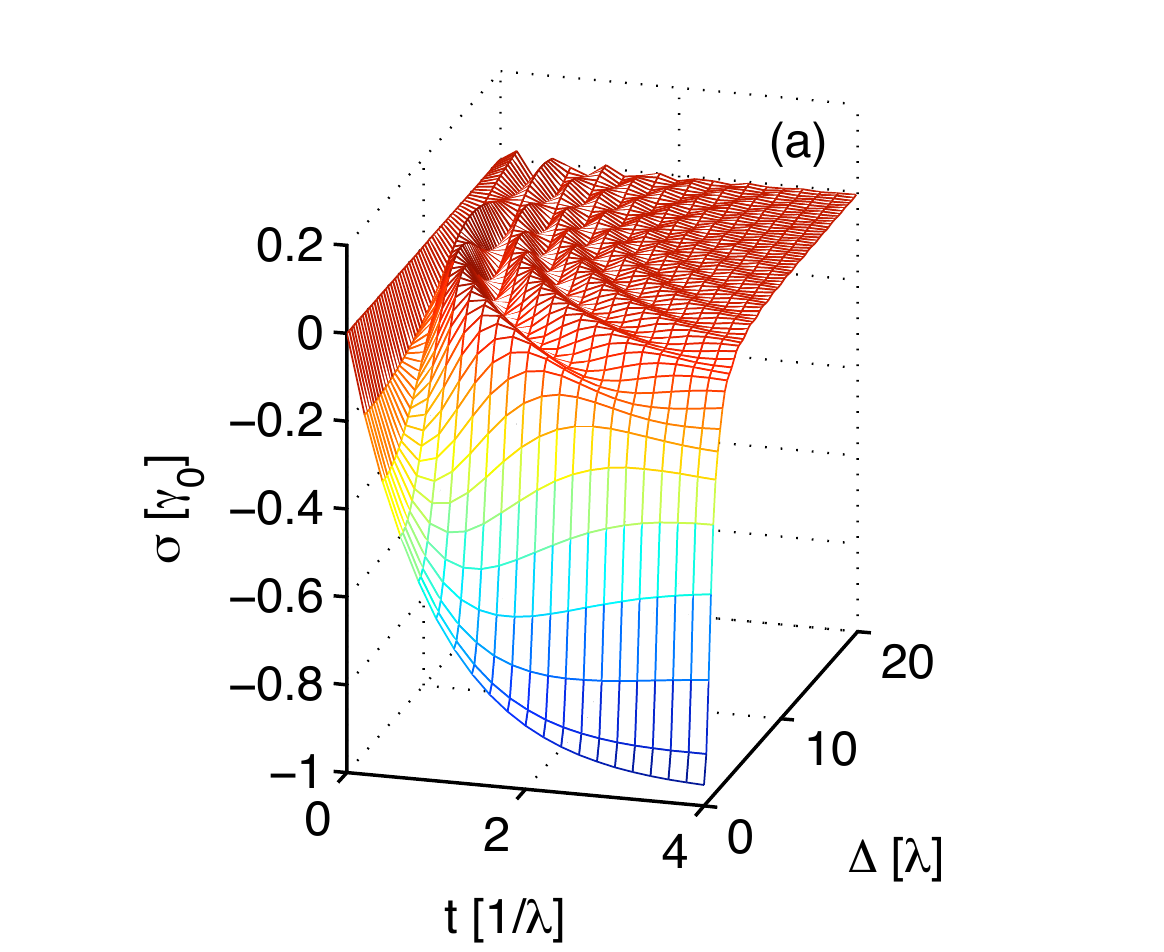}
\includegraphics[scale=0.34]{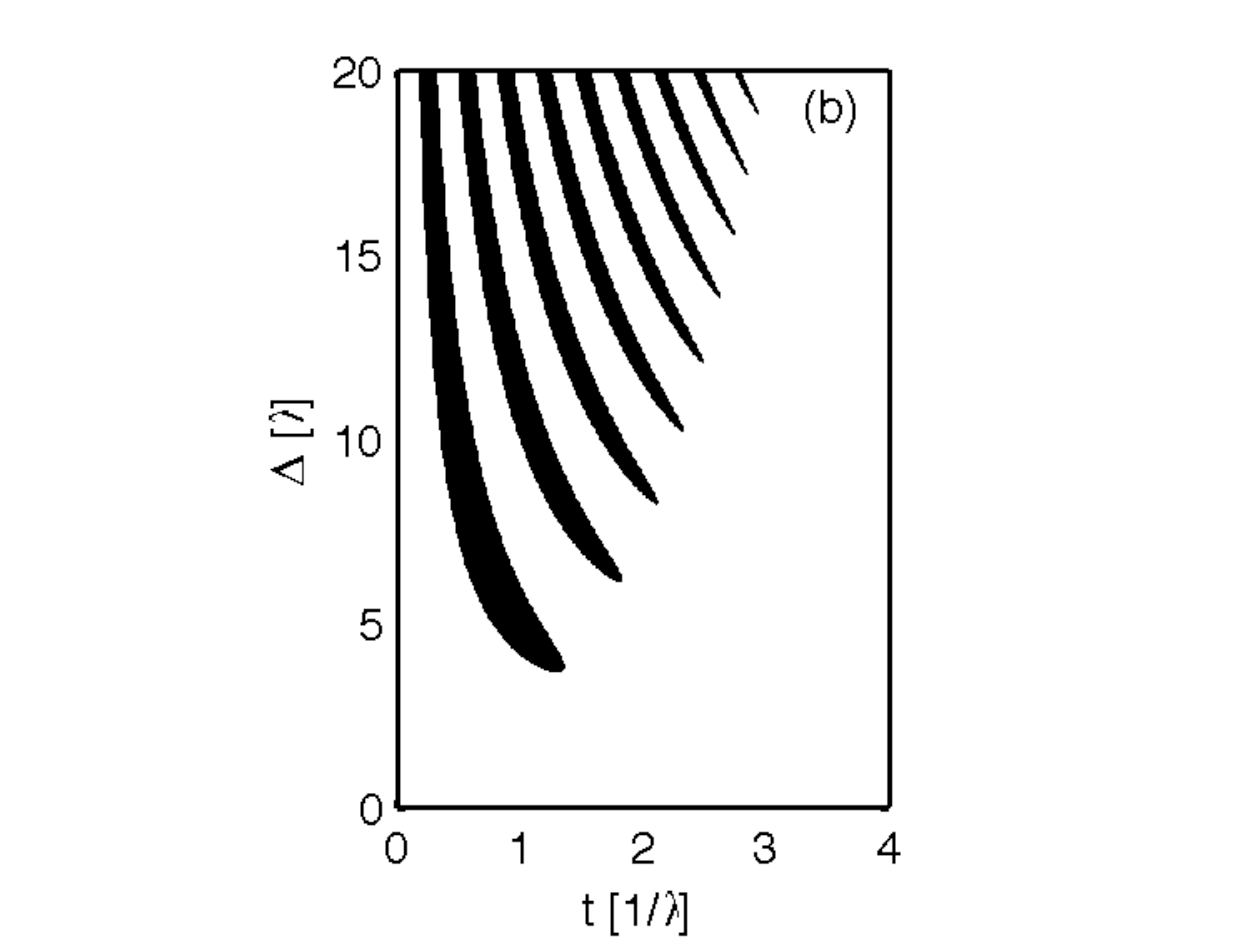}
\caption{(Color online) (a) The rate of change $\sigma$ of the
trace distance as a function of time $t$ and detuning $\Delta$ for
the initial pair of states $\rho_1(0)=|+\rangle\langle +|$ and
$\rho_2(0)=|-\rangle\langle -|$. (b) The black regions represent
the regions in which $\sigma$ is positive. \label{sigma-twolevel}}
\end{center}
\end{figure}

By construction, we have ${\mathcal{N}}(\Phi)=0$ for all processes
which have the divisibility property (\ref{DIVISIBILITY}). In the
following we discuss two simple examples for which our measure of
non-Markovianity is greater than zero. The aim is to illustrate
how to determine this quantity in specific cases and how
non-Markovianity is related to a violation of the divisibility
property and to the emergence of negative rates in master
equations of the structure (\ref{QMEQ-2}).

The first example describes a two-level system with excited state
$|+\rangle$ and ground state $|-\rangle$ which interacts with a
reservoir of field modes. The exact interaction picture master
equation \cite{BREUER2007} describing the dynamics of the density
matrix is of form of Eq.~(\ref{QMEQ-2}) with the generator
(\ref{TCL-GENERATOR}), where $H(t)=0$ and we have only a single
time-independent Lindblad operator $A=\sigma_-$ and a
time-dependent rate $\gamma(t)$. The function $\gamma(t)$ is
determined by the spectral density $J(\omega)$ of the reservoir.
We investigate the case of a Lorentzian spectral density
$J(\omega)=\gamma_0\lambda^2/2\pi[(\omega_0-\Delta-\omega)^2+\lambda^2]$,
the center of which is detuned from the transition frequency
$\omega_0$ of the two-level system by an amount $\Delta$, and work
in the weak coupling limit $\gamma_0/\lambda = 0.01$ (damped
Jaynes-Cummings model). For sufficiently large detunings $\Delta$,
the function $\gamma(t)$ then describes an exponentially damped
oscillation and takes on negative values within certain intervals
of time corresponding to a revival of the coherence in the system
\cite{BREUER2007,Piilo2007}. We emphasize that this does not imply
a violation of the complete positivity of the corresponding
dynamical map $\Phi(t)$ because the necessary and sufficient
condition for the complete positivity of $\Phi(t)$ is given by
$\Gamma(t)\equiv\int_0^t dt' \gamma(t') \geq 0$, which is indeed
satisfied here.

However, the trace distance increases for those $t$ for which
$\gamma(t)<0$, i.e., we have $\sigma(t,\rho_{1,2}(0))>0$ for these
times. This point is illustrated in Fig.~\ref{sigma-twolevel}
which shows $\sigma$ as a function of time $t$ and detuning
$\Delta$ for the pure initial states $\rho_1(0)=|+\rangle\langle
+|$ and $\rho_2(0)=|-\rangle\langle -|$. For these initial states
one finds the simple expression
\begin{equation}
 \sigma(t,\rho_{1,2}(0))=-\gamma(t)\exp[-\Gamma(t)],
\end{equation}
which shows that a positive $\sigma$ and an increase of the trace
distance is linked to a negative rate in the master equation.
Thus, the appearance of negative rates signifies a violation of
the divisibility property (\ref{DIVISIBILITY}) and a flow of
information from the environment back to the system.

\begin{figure}[htb]
\begin{center}
\includegraphics[width=0.8\linewidth]{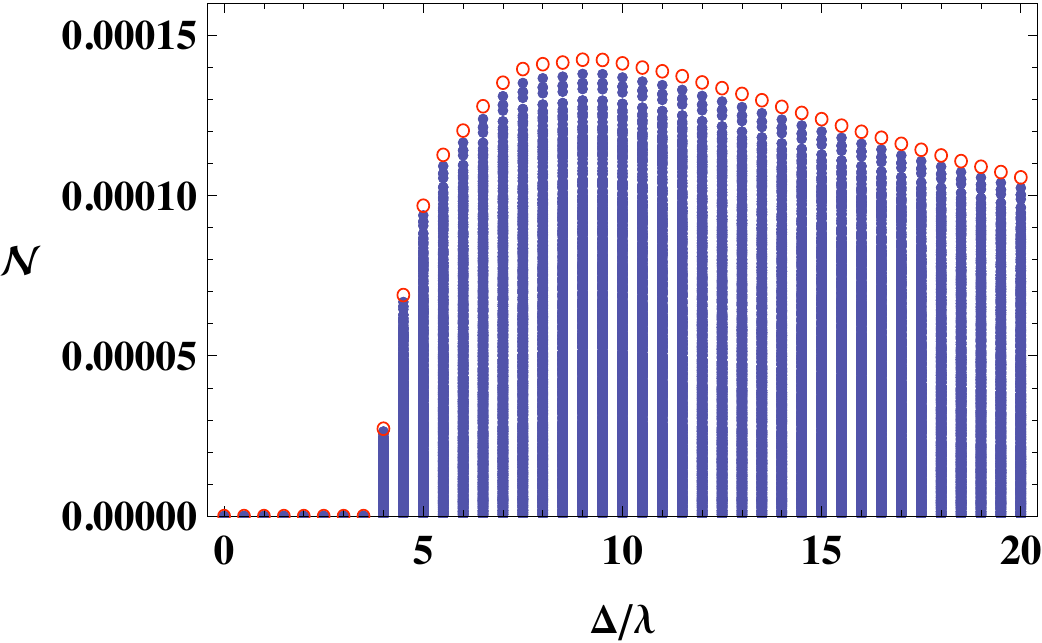}
\caption{(Color online) The non-Markovianity ${\mathcal{N}}(\Phi)$
for the damped Janes-Cummings model as a function of the detuning
$\Delta$. Blue dots: 1000 randomly drawn pairs of pure and mixed
initial states. Red circles: The initial pair
$\rho_1(0)=|+\rangle\langle +|$ and $\rho_2(0)=|-\rangle\langle
-|$ which leads to the maximum in Eq.~(\ref{MEASURE-1}).
\label{max-twolevel}}
\end{center}
\end{figure}

The maximization over the pair of initial states $\rho_{1,2}(0)$
in expression (\ref{MEASURE-1}) can be performed by drawing a
sufficiently large sample of random pairs of initial states. The
results are shown in Fig.~\ref{max-twolevel} and provide strong
numerical evidence that the maximum is attained for the initial
states $\rho_1(0)=|+\rangle\langle +|$ and
$\rho_2(0)=|-\rangle\langle -|$. This result could have been
expected since $\rho_2(0)$ represents the invariant state and
$\rho_1(0)$ has the largest distance to this state. According to
Fig.~\ref{max-twolevel} ${\mathcal{N}}(\Phi)$ exhibits a
non-monotonic behavior: The increase of the number of intervals in
which $\sigma > 0$ is overcompensated for large $\Delta$ by the
decrease of the size of $\sigma$ in these intervals.

For the previous example the non-Markovianity
${\mathcal{N}}(\Phi)$ was found to take on finite, positive
values. Our second example represents a rather extreme case,
demonstrating that there are also processes for which
${\mathcal{N}}(\Phi)$ is infinite. We consider a central spin with
Pauli operator $\bm{\sigma}$ which interacts with a bath of $N$
spins with Pauli operators $\bm{\sigma}^{(k)}$ through the
Hamiltonian $H=A\sum_{k=1}^N \sigma_z\sigma_z^{(k)}$, where $A$ is
a coupling constant. This simple model can easily be solved
exactly \cite{BREUER2004b}. Assuming the initial state of the bath
to be a maximally mixed state, one finds that the populations of
the density matrix of the central spin stay constant in time,
while the coherences are multiplied by the factor
$f(t)=\cos^N(2At)$. This leads to a simple formula for the trace
distance of the states $\rho_1(t)$ and $\rho_2(t)$,
\begin{equation}
 D(\rho_1(t),\rho_2(t)) = \sqrt{a^2+f^2(t)|b|^2},
\end{equation}
where $a=\rho^{++}_1(0)-\rho^{++}_2(0)$ denotes the difference of
the populations, and $b=\rho^{+-}_1(0)-\rho^{+-}_2(0)$ the
difference of the coherences of the two initial states. It follows
that the trace distance oscillates periodically between
$D_{\max}=\sqrt{a^2+|b|^2}$ and $D_{\min}=|a|$. This can be
interpreted as a periodic oscillation of the distinguishability of
the two states, i.e., as a periodic exchange of information
between the central spin and the spin bath.

The maximal growth of the trace distance occurs if one takes as
initial states the two eigenstates of $\sigma_x$, or any other
pair of states corresponding to antipodal points on the equator of
the Bloch sphere, such that $a=0$ and $|b|=1$. The trace distance
then oscillates periodically between the values $1$ and $0$. The
sum in Eq.~(\ref{MEASURE-2}) therefore diverges and we obtain
${\mathcal{N}}(\Phi)=+\infty$, which implies that a Markovian
approximation of the system dynamics is never possible. One can
formally write a master equation of the form (\ref{QMEQ-2}) with
$H=0$, a single Lindblad operator $A=\sigma_z$ and the rate
$\gamma(t)=AN\tan(2At)$, which shows again the connection between
the growth of the trace distance and the emergence of negative
rates.

Summarizing, we have constructed a measure ${\mathcal{N}}(\Phi)$
for the non-Markovianity of quantum processes in open systems. The
definition (\ref{MEASURE-1}) of the measure neither relies on any
specific representation or approximation of the dynamics, nor does
it presuppose the existence of a master equation or of invariant
states. The exact determination of the measure generally requires
solving the complete reduced system dynamics which could be a
difficult task for more complex problems. However, any observed
growth of the trace distance is a clear signature for
non-Markovian behavior and leads to a lower bound for
${\mathcal{N}}(\Phi)$. The measure for non-Markovianity introduced
here could therefore be useful also for the experimental
validation of theoretical models or approximation schemes. To
detect non-Markovianity experimentally one has to perform a state
tomography on different ensembles at different times in order to
decide whether or not the trace distance has increased. A great
advantage of the present approach is given by the fact that it
allows to plan experiments which test non-Markovianianity without
knowing anything about the properties of the environment or about
the structure of the system-environment interaction. Hence, we
think that our measure is a useful tool for the characterization
of non-Markovianity, both in theoretical descriptions and in
experiments.

\acknowledgments This work has been supported by the Academy of
Finland (Project Nos.~115982, 115682) and the Magnus Ehrnrooth
Foundation.

\end{document}